\documentclass[twocolumn,preprintnumbers,floatfix,prl,superscriptaddress]{revtex4}
\usepackage{graphicx}
\usepackage{graphicx}
\usepackage{dcolumn} 
\usepackage{bm}
\usepackage{epsfig}
\begin{document} 

\title{Half-metallic semiDirac point generated by quantum confinement \\
in TiO$_2$/VO$_2$ nanostructures}

\author{Victor Pardo}
\affiliation{Department of Physics,
  University of California, Davis, CA 95616
}
\affiliation{
Departamento de F\'{\i}sica Aplicada, Universidade
de Santiago de Compostela, E-15782 Santiago de Compostela,
Spain
}

\author{Warren E. Pickett}
\affiliation{Department of Physics,
  University of California, Davis, CA 95616
}


\begin{abstract}

Multilayer VO$_2$/TiO$_2$ nanostructures ($d^1$ - $d^0$ interfaces with no polar
discontinuity)
are studied with first principles density functional methods
including structural relaxation.
Quantum confinement of the {\it half metallic} VO$_2$ slab within insulating
TiO$_2$ produces an
unexpected and unprecendented two-dimensional new state, with a {\it semiDirac} point Fermi
surface: spinless charge
carriers are effective-mass like along one principal axis, but are
massless along the
other. Effects of interface imperfection are addressed.

\end{abstract}

\maketitle

VO$_2$ is a magnetic oxide that undergoes a metal-insulator transition \cite{vo2_mit} upon 
lowering the temperature through 340 K, accompanied of a symmetry-breaking structural 
transition from the high temperature metallic rutile phase \cite{vo2_rutile}. The 
insulating state takes place via a dimerization of the V-V chains \cite{vo2_dimer}.  
The origin of this metal-insulator transition is the focus of much recent theoretical 
activity and remains uncertain. It could be due to the formation of a Peierls 
state \cite{vo2_peierls1,renata}, or it could be 
driven by correlations \cite{mottI,vo2_corr},  or more likely may
have some mixed origin \cite{spin-P,vo2_peierls2}.
TiO$_2$ is isostructural (in one of its phases) and is a d$^0$ non-magnetic insulator 
that is very important industrially and is well understood.

The interface (IF) between a correlated insulator and a band insulator has been recognized as 
fertile ground for new behavior \cite{ohtomo,hwang}, and the LaTiO$_3$/SrTiO$_3$ 
(LTO/STO) IF involving the Mott insulator LTO has attracted much of the theoretical 
study to date \cite{hamann,spaldin,pentcheva2007}.
For IFs between band insulators, LaAlO$_3$/SrTiO$_3$ (LAO/STO) has received a great deal
of attention \cite{pentcheva2006,siemons,freeman,pentcheva2008,willmott}.
In both cases there is a polar discontinuity across the IF, and this
aspect has been expected to dominate the resulting behavior, and lead to unexpected phenomena.

The (001) VO$_2$/TiO$_2$ IF has been studied by photoemission spectroscopy (PES) \cite{pes_prb},
which found the IF is insulating when the VO$_2$ substrate is insulating.  PES also has
uncovered \cite{okazaki} spectral weight transfer in VO$_2$/TiO$_2$ thin films 
indicating strong correlation
effects even for conducting VO$_2$.  Much of the focus on this nanostructure has been on tuning the 
VO$_2$ metal-insulator transition temperature \cite{if_apl}, because of its potential 
technological applications. It has been found that a minimum thickness 
of 5 nm of VO$_2$ is needed to sustain a metal-insulator transition, for thinner VO$_2$ 
layers the transition no longer occurs \cite{vo2_jap} (the VO$_2$ layer remains conducting).
The explanation is that the insulating
state requires a collective structural dimerization along the rutile $c$-axis that is inhibited
by confinement for thinner layers.
Since the IF is not polar and the lattice mismatch is small (1 \%), structural relaxation
is not expected to be severe.  
Any unusual behavior of this multilayer (ML) will require different microscopic
mechanisms than have been uncovered before.

In this paper we present a theoretical study of the electronic behavior of the 
ML nanostructures 
(TiO$_2$)$_n$/(VO$_2$)$_m$, denoted ($n/m$), looking in particular at the evolution 
of the conduction and magnetic 
properties with VO$_2$ layer thickness.  Since VO$_2$ is the component that is potentially
conducting, we focus on thin VO$_2$ layers which will incorporate any consequences of
quantum confinement.  The properties are found to depend strongly on layer thickness
and the effects of quantum confinement at small thicknesses give rise to a new electronic
state for certain MLs.

\begin{figure}[ht]
\begin{center}
\includegraphics[width=\columnwidth,draft=false]{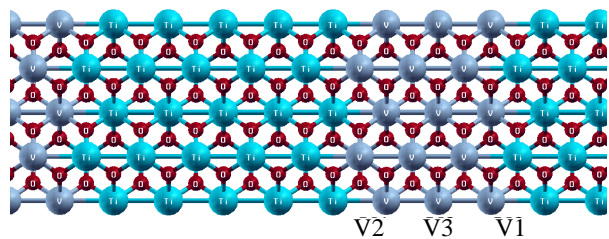}
\caption{(Color online) Structure of the 5/3 TiO$_2$/VO$_2$ supercell corresponding to 
growth along the (001) axis, which is the 
metal chain direction of the rutile structure.  V1, V2, V3 label the V ion sites beginning
from the one nearest to TiO$_2$. (Due to a symmetry with respect to the center of the 
VO$_2$ slab, the V layers are V1-V2-V3-V3-V2-V1.) }\label{struct}
\end{center}
\end{figure}

\begin{figure}[ht]
\begin{center}
\includegraphics[width=\columnwidth,draft=false]{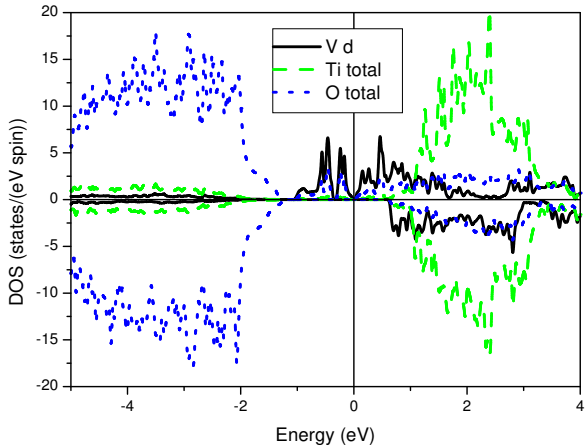}
\caption{(Color online) Total density of states in the 5/3 multilayer, showing the 
location of the V and Ti bands relative to those of O (which are unpolarized).  
The V occupied majority spin bands (plotted
upward) lie within a 2 eV gap in the minority spin.  }\label{DOS}
\end{center}
\end{figure}

Starting from an average rutile structure of ($n/m$) MLs, we performed 
volume and $c/a$ optimization and an internal 
relaxation of the atomic positions of all the atoms. 
The main modification is lattice strain along the $c$ axis, interpolating between the slightly
different $c$ lattice constants.  
We have studied several MLs, varying both TiO$_2$ and VO$_2$ layer 
thicknesses, following identical procedures.
Our electronic structure calculations were  performed within density functional 
theory \cite{dft} using the all-electron, full potential code {\sc wien2k} \cite{wien}  
based on the augmented plane wave plus local orbital (APW+lo) basis set \cite{sjo}.
The exchange-correlation 
potential utilized to deal with possible strong correlation effects was the LSDA+U 
scheme \cite{sic1,sic2} including an on-site U and J (on-site Coulomb repulsion and exchange strengths) 
for the Ti and V $3d$ states.  The values U=3.4 eV, J= 0.7 eV have been used for both Ti and V to deal properly with 
correlations in this multilayered structure; these values are comparable (slightly smaller) than
what have been used for bulk VO$_2$ \cite{vo2_peierls2,tomczak,haverkort}.  In the paper spin-orbit
coupling has been neglected.

While we have studied a variety of $n/m$ (001) MLs, we focus primarily on the (5/3) ML
(1.5 nm TiO$_2$, 0.9 nm 
of VO$_2$).  The structure, and the identification of the three distinct V
sites, is shown in Fig. \ref{struct}.  In terms of distance from the TiO$_2$
layer, the V ions are labeled V1, V2, V3.  
The tetragonal symmetry of the rutile structure has been retained in the $x-y$ plane.

\begin{figure*}[ht]
\begin{center}
\includegraphics[width=16cm,draft=false]{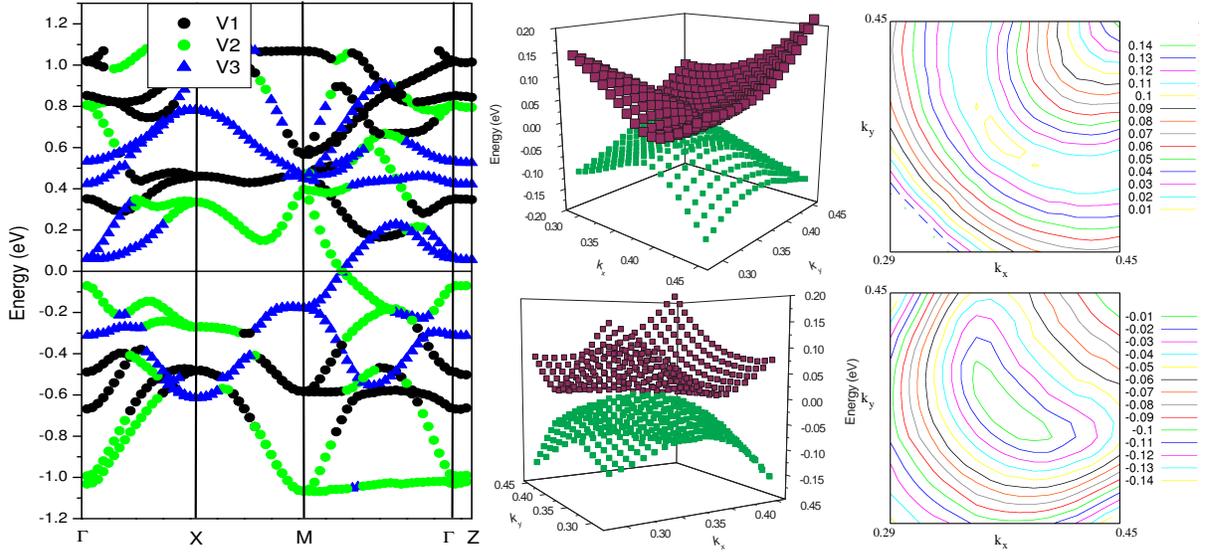}
\caption{(Color online) Left panel: Blow-up of the band structure around the Fermi level showing in different colors the biggest character of each band. Notice that the two bands crossing at the Fermi level have character from the two most inner V atoms.
Note the semiDirac point along the (110) direction where bands cross precisely at the Fermi level. 
Middle panel: Two different views of the same `surface' plot of the two bands 
that cross the Fermi level, centered around the semiDirac point. The valence and 
conduction bands cross at a single point.  The linear dispersion can be seen in the
upper plot (upper left and lower right); the quadratic dispersion is clear in the 
lower panel, where the flatness of the conduction band is also clear. Right panel: 
contour plots at constant energy (in eV, relative to the Fermi level) of the 
valence band (below), and the flat conduction band (above) that leads to large
M-centered Fermi surfaces for electron doping.}\label{bs}
\end{center}
\end{figure*}

V $3d$ bands (Fig. \ref{DOS}) dominate the spectrum close 
to the Fermi level (E$_F$), and only three TiO$_2$ cells are required to give negligible
$k_z$ dispersion and thus confine the $3d$ states to a two dimensional (2D) system. 
FM alignment of the spins is preferred, and half metallicity results.
Enlargement of the density of states (DOS) (not shown) reveals vanishing of the DOS
precisely at E$_F$, with no V1 participation just below E$_F$.
This curious vanishing of the DOS reflects a zero gap semiconductor involving V2 and V3 ion states.

The majority spin band structure along high symmetry directions shown in Fig. \ref{bs}
clarifies an unexpected and unprecendented electronic state.
Two bands cross the Fermi level at a single point along the zone diagonal at 
the point $k_{sD}$=($\pm 0.37,\pm 0.37$)$\pi/a$
(the precise position of $k_{sD}$ along the (1,1) 
direction depends on the 
value of U).  Inspection throughout the zone confirms that 
this Fermi surface crossing is a single point (rather, four symmetry related points), 
as is the Dirac point in graphene \cite{graphene_nat05}.
This single point determines the Fermi
energy, again as in graphene \cite{katsnelson}.  The crossing
of the bands precisely at E$_F$ is therefore not accidental, rather it is topologically
determined: there are exactly six filled bands below this point, containing the 
majority spin electrons of each of the six V ions in the cell. 

These two bands crossing E$_F$ involve separately V2 and V3 ion $3d$ states, as is
illustrated with the color coded fatbands in Fig. \ref{bs}.  With no contribution 
from the IF ion V1, the dominance of the interior ions V2 and V3 
identifies this as a {\it quantum confinement effect} rather than an IF phenomenon. 
Additional VO$_2$ layers, which relieve the confinement effects,
add more bands and introduce a Fermi surface.
In Fig. \ref{bs} we provide a surface plot of 
these two band energies in a small region in k-space centered on the band crossing 
point $k_{sD}$.  This state results only
after the ion positions are relaxed, and arises due to band reordering at $k$=0 that 
occurs during the relaxation.  The surface plot reveals
yet another peculiarity: while the dispersion is linear along the (1,1) direction
as is clear from the band plot, the dispersion is {\it quadratic} perpendicular to
the diagonal; the gap opens due to the loss of
symmetry of the two bands off
the diagonal $k_x = k_y$, and does so quadratically.  To differentiate this point 
from the graphene Dirac point, we refer
to it as the (half metallic) semiDirac ($sD$) point. 
The corresponding mass tensor shows
extreme anisotropy (zero to normal values), as does the velocity (1.5 $\times$ 
10$^7$ cm/s to zero) \cite{graphene_nat05}. This very strongly anisotropic dispersion, between
extremes (normal values, to zero)
will give rise to peculiar transport and thermodynamic properties, which will
be reported separately.

The constant energy surfaces for both electrons and holes are plotted in 
the right panel of Fig. \ref{bs} for low energies in a small region around $k_{sD}$.  
The conduction band has a flatness that opens a path for a Fermi surface to develop 
as a ring around the M=($\pi,\pi$) point at very low electron doping. 
The valence band shows iso-energetic 
contours with an elliptical shape, with the longer axis perpendicular to the
zone diagonal.

\begin{figure}[ht]
\begin{center}
\includegraphics[width=\columnwidth,draft=false]{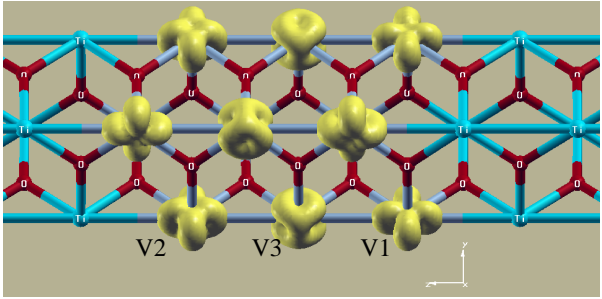}
\caption{(Color online) Spin density plot, isosurface at 0.15 e/\AA$^3$. The particular orbital ordering in a spin-aligned configuration is shown. V1 and V2 have one electron in a d$_{\parallel}$ and V3 is in a d$_{\perp}$ orbital.}\label{rho}
\end{center}
\end{figure}

In bulk VO$_2$ (V$^{4+}$: d$^1$ cations)  
the distortion from cubic symmetry of the VO$_6$ octahedron introduces a crystal field 
(actually, ligand field)
that lifts the degeneracy of the t$_{2g}$ orbitals, splitting them into a d$_{\parallel}$ singlet and 
two d$_\perp$ orbitals (using Goodenough's notation \cite{vo2_peierls1}). 
The orbital ordering that arises in this 5/3 ML is illustrated 
by the spin density isosurfaces in
Fig. \ref{rho}.
The V ions (V1 and V2) that terminate a V-V-V chain have an occupied d$_{\parallel}$ orbital, whereas the 
chain-center V3 ion has an occupied
orbital of d$_\perp$ of combined d$_{xz}$ and d$_{yz}$ character. This orbital ordering is dependent 
on the magnetic ordering; when the spins along the $c$ direction are antialigned 
($\uparrow \downarrow \uparrow$)  
the d$_{\parallel}$ orbital becomes occupied in all sites.  

{\it Influence of VO$_2$ slab thickness.}  A metal-insulator 
transition is observed \cite{vo2_jap} for VO$_2$ thicknesses above 5 nm (approximately 15 layers), 
but experimental information on crystalline samples with smaller thickness is sparse \cite{lam}.
Our calculations show that the system has an insulating ground state for two  
layers of VO$_2$, where spin antialignment is favored.
However, for a thickness of approximately one nm (3 layers), the material is in the intermediate
 zero-gap semiDirac state described above, on the brink of metallicity. 
Thicker VO$_2$ layers (four or more) become half metallic, a property that
is much sought in oxide nanostructures because of its potentially enormous 
technological applications in spintronic devices.

{\it Influence of magnetic alignment.}
We have studied antialignment of the moments  (ferrimagnetism)
along the (001) V-V chains. Such AF coupling is energetically unfavorable in almost all cases.
Interestingly, such antialignment 
changes the orbital ordering: in the 5/3 multilayer (the semiDirac point system when FM), 
flipping the spin of the intermediate V ion (V2) results in all V ions having an occupied
d$_{\parallel}$ orbital, because 
the $\sigma$-bond along the z-axis 
between neighboring d$_{\parallel}$ orbitals favors AF coupling. 

{\it Role of V-Ti exchange disorder.}  States that are very sensitive to 
disorder are less likely to have
importance in applications, since thin film growth does not result in perfectly ordered materials,
so we have begun study of the effect of V/Ti exchange near the IF.
We find that the most unexpected feature, {\it i.e.} the development of a half metallic 
semiDirac point for three VO$_2$ layers,  
is robust with respect to two types of 
ion exchange that do not change the electron count, {\it i.e.} no doping. The first type 
was the interchange of V1 with Ti across the IF, which is a typical defect in growth. 
If we label the Ti sites across the multilayer as Ti1/Ti2/Ti3/Ti4/Ti5/Ti5/Ti4/Ti3/Ti2/Ti1, 
this first type of disorder corresponds to the interchange of V1 and Ti1, corresponding to a non-abrupt IF.
The second  type 
is to interchange V1 with Ti2, which are neighbors along the cation chain. 
In both cases a semiDirac point persists in spite of changes of the band structure,
and confirms that it is the V2 and V3 ions are produce the active bands. 

To address the robustness of this unusual property in the band structure for the 3-layer 
VO$_2$ system, we have varied the thickness of the confining TiO$_2$ layer.  Reduction of the 
TiO$_2$ slab thickness 
to just three layers changes slightly the bands and thereby the
position of the crossing point in the Brillouin zone, 
but still gives negligible dispersion along the z-axis, {\it i.e.} the behavior is still
2D. The semiDirac point is also robust with
respect to the strength of correlation effects: the semiDirac point 
varies along the diagonal from (0.3,0.3) to (0.4,0.4) 
depending on both the choice of U on the V ions (for reasonable values, above 2 eV) and TiO$_2$ thickness.

The finding that quantum confinement, together with specific orbital occupation and perhaps
important symmetries, in oxide multilayers can produce a semiDirac point
at the crossover between insulating and conducting behavior introduces a novel feature in
the physics of oxide heterostructures: a polar discontinuity is not required to produce
unexpected and unprecedented electronic states in these systems.  The transport behavior,
and the changes with doping, for systems with a semiDirac point will be addressed in following 
papers, as will the complicating effects of spin-orbit coupling.  
We note that oxide nanostructures are mechanically more robust than graphene, which 
could make patterning of such multilayers possible.


This project was supported by DOE grant DE-FG02-04ER46111 and through interactions with
the Predictive Capability for Strongly Correlated Systems team of the Computational
Materials Science Network and a collaboration supported by a Bavaria-California Technology
grant.  V.P. acknowledges financial support from Xunta de Galicia (Human Resources Program).


\end{document}